\documentclass[amsmath,final,3p,times,twocolumn]{elsarticle}
\usepackage[utf8]{inputenc}
\usepackage{amsfonts}
\usepackage{amssymb}
\usepackage{rotating} 
\usepackage{graphicx} 
\usepackage{epsfig}
\usepackage{multirow}
\usepackage{bm}
\usepackage{tabularx}
\usepackage{xcolor}
\usepackage{hyperref}
\usepackage{subcaption}

\usepackage[toc,page]{appendix}
\usepackage{amsmath}
\usepackage{cleveref}

\newcommand{\diracslash}[1]{#1\llap{/\kern2pt}}

\newcommand{\be}{\begin{equation}}
	\newcommand{\ee}{\end{equation}}
\newcommand{\bea}{\begin{eqnarray}}
	\newcommand{\eea}{\end{eqnarray}}
\newcommand{\ba}[1]{\begin{array}{#1}}
	\newcommand{\ea}{\end{array}}

\newcommand{\bt}{\begin{tabular}}
	\newcommand{\et}{\end{tabular}}

\newcommand{\beas}{\begin{eqnarray*}}
	\newcommand{\eeas}{\end{eqnarray*}}

\begin{document}
	\begin{frontmatter}
		
		\title{Strange quark stars in modified vector MIT bag model: role of $\rho$ and $\phi$ mesons} 
		\author{Mukul Wadhwa}
		\ead{mukulwadhwa99@gmail.com}
		\author{Manisha Kumari}
		\ead{maniyadav93@gmail.com}
		\author{Arvind Kumar}
		\ead{kumara@nitj.ac.in}
		
		\affiliation{organization={
				Department of Physics, Dr. B R Ambedkar National Institute of Technology Jalandhar},
			city={Jalandhar},
			postcode={144008}, 
			state={Punjab},
			country={India}}
		


		\def\be{\begin{equation}}
			\def\ee{\end{equation}}
		\def\bearr{\begin{eqnarray}}
			\def\eearr{\end{eqnarray}}
		\def\zbf#1{{\bf {#1}}}
		\def\bfm#1{\mbox{\boldmath $#1$}}
		\def\hf{\frac{1}{2}}
		\def\kp{\zbf k+\frac{\zbf q}{2}}
		\def\km{-\zbf k+\frac{\zbf q}{2}}
		\def\hwo{\hat\omega_1}
		\def\hwt{\hat\omega_2}

		\begin{abstract}
			In the present work, we study the properties of strange quark stars (SQSs) using the vector MIT bag model with modification in vector channels. 
			Unlike recent studies which only consider interactions through $\omega$ mesons, we analyze the possibility of $\rho$ and $\phi$ vector channels. We consider two types of higher order non-linear self-interaction terms for the vector mesons. With these modifications,  we computed the equation of state (EoS) and mass-radius of strange stars for different values of vector coupling strength. 
		 Considerations of $\rho$ and $\phi$ vector mesons along with $\omega$, as well as an increase in the strength of vector coupling $g_v$, enhance the mass and radius of SQSs.
			For two kind of non-linear self-interactions of vector mesons
			considered in the present calculations, we
			observe the SQSs with maximum mass $2.48$ and $2.42  M_{\odot}$ for the vector coupling $g_v = 3$. Corresponding radii of these SQSs are
			found to be  $12.27$ and $12.18$ km, respectively. 
			We also calculate the tidal deformability parameter $\Lambda$, the Love number $k_2$
			and the gravitational redshift of SQSs. 	 The tidal deformability 
			parameter $\Lambda$ is observed to increase with $g_v$, with appreciable effect for low mass stars.
		\end{abstract} 
		\begin{keyword}
			Strange quark stars, MIT bag model, Vector interactions
		\end{keyword}
		
	\end{frontmatter}
	\section{\label{intro}Introduction}
	In the framework of quantum chromodynamics (QCD), which is the theory of strong force governing the interactions between quarks and gluons, different phases of matter are depicted through the QCD phase diagram in the plane of temperature and baryonic density. At high temperature and low baryon density, such as conditions during the early universe, the phase of matter having free quarks and gluons known as quark-gluon plasma (QGP) may exist. 
	At zero or very low temperatures and high baryon density, as found at the cores of massive objects such as neutron stars, there may be a possibility of deconfined quark matter. The core of these massive objects is subjected to intense gravitational binding energy and high baryonic density, at such extreme conditions, quark matter can become energetically more favorable over hadronic matter. According to the strange matter hypothesis, the strange matter may be the true ground state of strongly interacting matter \cite{Bodmer:1971we,Terazawa:1979HT,Witten:1984rs}. Strange matter is a kind of quark matter that consists of up, and down along with strange quarks. It also suggests that when the core of a neutron star undergoes a phase transition from hadronic to quark phase, the entire core can convert into a strange quark star (SQS) \cite{Olinto:1986je}. 
	
	Different non-perturbative models, for example, MIT bag model and its various extension \cite{Chodos:1974je, Chodos:1974pn, DeGrand:1975cf, Gomes:2018bpw, Franzon:2016urz, Lopes:2020btp}, quark mass density dependent (QMDD) model \cite{Lugones:2022upj, Benvenuto:1998tx, Chu:2012rd, Peng:1999gh, Wen:2005uf}, chiral SU(3) quark mean field model \cite{Kumari:2021tik, Kumari:2020mci}, Nambu Jona Laisinio (NJL)  \cite{Li:2019ztm, Wang:2019gam, Buballa:2003qv, Hatsuda:1994pi, Lenzi:2010mz}, quasi-particle model \cite{Li:2019akk, Zhang:2021qhl, Chu:2023rty, Pal:2024nza} etc. have been used to study the equation of state (EoS) of high density matter and analyze the properties of strange quark matter and SQSs.
	The observation of massive pulsars PSR  J1614-2230  with mass $1.97 \pm 0.04$  $M_{\odot}$ \cite{Demorest:2010bx}, PSR J0348+0432 having mass $2.01 \pm 0.04$ $M_{\odot}$ \cite{Antoniadis:2013pzd}, PSR J2215+5135 with mass $2.27^{+0.17}_{-0.15} M_{\odot}$ \cite{Linares:2018ppq} and PSR J0740+6620 with mass $2.14^{+0.20}_{-0.18} M_{\odot}$ \cite{NANOGrav:2019jur} have ruled out the possibilities of soft EoS and triggered research interest in direction to improve the phenomenological approaches so as produce stiffed EoS and to confront the theoretical studies with astronomical observations. The detection of gravitational waves by LIGO/Virgo collaboration from binary star merger for GW170817 \cite{LIGOScientific:2017vwq, LIGOScientific:2018cki} and GW190814 \cite{LIGOScientific:2020zkf} events also put stringent constraints on EoS through observation on masses and tidal deformability parameter of compact objects.
	Neutron Star Interior Composition Explorer (NICER)
	of NASA also put constraints on mass-radii of these compact objects \cite{Riley:2021pdl}.
	
	In our present work we shall use the vector MIT bag model with modified vector channel to explore the properties of SQSs.
	Vector MIT bag model is used to describe the properties of quarks and gluons confined in a finite region called \textquotedblleft bag" by a quantity called bag pressure, which accounts for the strong force that confines the quark inside hadrons \cite{Chodos:1974je, Chodos:1974pn, DeGrand:1975cf}.
	In the original MIT bag model, quarks were treated as non-interacting, however, the different versions of model which consider the interaction among the confined quarks have been proposed \cite{Fraga:2001id,Alford:2004pf}.
	With these interactions being considered, the non-ideal bag model provides a more realistic description of quark matter in extreme conditions, such as those found in compact stars.
	Another approach that was used to include interactions among quarks was coupling them to a mediating vector field and mode is known as the vector MIT bag model. Here, quarks still remain confined inside the bag but they interact through the vector field. Having repulsive vector interactions makes the EOS stiffer, which further leads to an increase in possible maximum mass \cite{Gomes:2018bpw,Franzon:2016urz}. In recent versions of vector MIT bag model, a self-interacting term was added to mimic the contribution from the Dirac sea \cite{Lopes:2020btp}.
	The Dirac sea contributes to the overall dynamics of the system and it represents the infinite sea of negative energy states for fermions \cite{Lopes:2020btp}.
	Including this self-interaction term for vector mesons allows to enhance the model's capability to add complexities of quark interactions. Furthermore, to maintain the thermodynamic consistency of the model, mesonic mass was also introduced \cite{Lopes:2020btp}. 
	The MIT bag model with density dependent bag constant along with vector interactions has also been used to study the properties of quark stars \cite{Podder:2023dey, Ju:2024tvv} and hybrid stars \cite{Podder:2023dey, Kumar:2023lhv, Sen:2021cgl, Sen:2022lig, Pal:2023quk}.
	In Ref. \cite{Pal:2023dlv} the thermodynamic consistency of MIT bag model with  density or chemical potential dependent bag constant is explored.
	
	The non-linear self interaction terms used in vector MIT bag model were inspired from the relativistic mean field models \cite{Furnstahl:1996zm, Papazoglou:1998vr}.
	However, as per our knowledge, the vector MIT models which have been used so far in the literature considered only non-strange vector-isoscalar channel, i.e., the $\omega$ vector meson \cite{Lopes:2020btp}. 
	The vector-isovector $\rho$ meson and the strange vector-isoscalar $\phi$ have not been included yet in the MIT bag model, even without the self-interactions.
	In the relativistic mean field model studies of hadronic matter and compact stars, the $\rho$ meson is considered important since the medium has finite isospin asymmetry \cite{Papazoglou:1998vr}, whereas the presence of hyperons (having strange quark) necessitates the $\phi$ meson \cite{Papazoglou:1997uw}. The strange quark stars, which are subject of our present study, have finite isospin asymmetry due to unequal number density of $u$ and $d$ quarks and also have $s$ quark of-course, and therefore, investigating their properties using vector MIT bag model, considering  $\rho$ and $\phi$ mesons (in addition to $\omega$), can have important consequences.
	Therefore, in the present work, we shall investigate the mass-radii and tidal-deformability parameter for SQSs  using vector MIT bag model with modified vector channel considering $\omega, \rho$ and $\phi$ mesons.
	We shall take into account higher order non-linear self interactions of vector mesons considering  two different forms for corresponding Lagrangian densities \cite{Dexheimer:2008ax,Cruz-Camacho:2024odu}. While taking into account these different vector mesons and  forms for self interactions, the thermodynamic consistency for the stability of strange quark stars will be ensured.
	
	The present article is organized as follows:
	In Sec. \ref{sec:vbag_model}
	we present the details of vector MIT bag model considering different forms for non-linear self interactions of $\omega, \rho$ and $\phi$ mesons.
	Tolman-Oppenheimer-Volkov (TOV) equations
	used to calculate the mass-radius relations will be presented in  Sec. \ref{Sec_tov}.
	In Sec. \ref{sec:results}, we present the  results and discussion on the properties of SQSs. Finally, summary and conclusion of the present work is given in Sec. \ref{sec:summary}.

	\section{Vector MIT bag model with modified vector channel} \label{sec:vbag_model}
	As discussed above, in the present work we shall use the vector MIT bag model, considering the interactions among  quarks through the exchange of $\omega, \rho$ and $\phi$ mesons. The Lagrangian density of such version of vector MIT bag model  can be written as
	\begin{align}
		\mathcal{L} &= \sum_{i=u,d,s}\{ \bar{\psi}_i  [ \gamma^{\mu} \left(i\partial_\mu - \left(g^{i}_{\omega} \omega + g^{i}_{\rho} \rho +
		g^{i}_{\phi} \phi \right)
		\right) - m_i ]\psi_i \nonumber\\
		&- B \}\Theta(\bar{\psi_i}\psi_i) + \frac{1}{2}\left(m_\omega^2 \omega^2 + m_\rho^2 \rho^2 +m_\phi^2 \phi^2\right) \nonumber \\ 
		&
		+ \mathcal{L}_{vec}^{Non} + \sum_{l=e,\mu} \bar{\psi}_l [i\gamma^\mu\partial_\mu -m_l]\psi_l. 
		\label{Eq_Lag_MIT1}
	\end{align}
	In the above equation, $B$ is the bag pressure which confine quarks within certain region and $\Theta(\bar{\psi_i}\psi_i)$ is the heavy-side function. 
	From the Dirac term of above Lagrangian density, the energy eigen value of a given quark, modified due to vector fields $\omega, \rho$ and $\phi$, is written as
	\begin{equation}
		e_i = \sqrt{k^2 + m_i^2} +
		g^{i}_{\omega} \omega + g^{i}_{\rho} \rho +
		g^{i}_{\phi} \phi.
		\label{Eq_eig_quark1} 
	\end{equation}
	In Eq. (\ref{Eq_Lag_MIT1}) the terms having mass squared expressions, $m_j^2, j = \omega,\rho,\phi$ are known as mass terms for the vector meson fields.
	Also, $g^{i}_{\omega}, g^{i}_{\rho}$ and
	$g^{i}_{\phi}$ represent the coupling of quarks with vector mesons.
	The term $\mathcal{L}_{vec}^{Non}$ considers the contribution from higher order non-linear self-interactions of vector mesons. Various forms of the fourth-order self-interaction term of the vector meson have been considered in Ref.~\cite{Dexheimer:2008ax}.
	We shall consider following two different forms for the
	higher order non-linear interactions, 
	\begin{align}
		\mathcal{L}_{vec-I}^{Non} &= 2 c_4 \text{Tr} \left(g^M V^4\right),
		\label{Eq_Lag_non1}\\
		\mathcal{L}_{vec-II}^{Non} &=  c_4 \left[\text{Tr} \left(g^M V^2 \right) \right]^{2}.
		\label{Eq_Lag_non2}
	\end{align} 
	The parameter $c_4$ is a dimensionless quantity and is set to unity in our calculations \cite{Lopes:2020btp}.
	The coupling matrix $g^M$ introduced in the above equations and vector meson matrix $V$ are defined through the diagonal matrices $
	g^M = \text{diag}\left( g_\omega^u, g^u_\rho, g^s_\phi\right)$ and $V = \text{diag} \left(\frac{\omega+\rho}{\sqrt{2}},\frac{\omega-\rho}{\sqrt{2}},\phi\right)$, respectively.
	In the present calculations, we consider $g_\omega^u = g_\omega^d = g_\rho^{u} = -g_\rho^d = \frac{g_{\phi}^s}{\sqrt{2}} = g_v$ \cite{Wang:2002pza}. Also, the
	coupling constant of light $u$ and $d$
	quarks with strange vector field $\phi$ and the strange $s$ quark with non-strange vector fields $\omega$ and $\rho$ is assumed zero, i.e., 
	$g_\omega^s = g_\rho^{s} = g_\phi^u = g_\phi^{d} = 0$. 
	In Ref. \cite{Lopes:2020btp},  the self-interactions of form given by  
	Eq. (\ref{Eq_Lag_non1}) and considering only vector meson $\omega$ is considered
	in the vector MIT model.
Explicitly, 
Eqs. (\ref{Eq_Lag_non1}) and (\ref{Eq_Lag_non2}) are expanded to the following
\begin{align}
	\mathcal{L}_{vec-I}^{Non} &=  c_4 \left[g^{u}_{\omega}\left(\omega^4 + \rho^4 + 6 \omega^2 \rho^2\right) +
	2g^{s}_{\phi}  \phi^4 \right], 
	\label{Eq_Lag2_non1}\\
	\mathcal{L}_{vec-II}^{Non} &=  
	c_4 \left[g^{u}_{\omega}\left(\omega^2 +  \rho^2  \right) +
	g^{s}_{\phi}  \phi^2 \right]^2.
	\label{Eq_Lag2_non2}
\end{align}
One another possibility of higher order non-linear interaction is  to consider the Lagrangian density  of form $
\mathcal{L}_{vec-III}^{Non} =  \frac{c_4}{4} \left[\text{Tr} \left(g^M V \right) \right]^{4} = 
\frac{c_4}{4}   \left[\sqrt{2} g^{u}_{\omega} \omega  +
g^{s}_{\phi}  \phi \right]^4$ \cite{Dexheimer:2008ax, Kumari:2022jvq}.
However, in our calculations, this term is observed to produce a soft EoS with mass less
than $2 M_{\odot}$.
The equations of motion for the vector fields $\omega, \rho$ and $\phi$ are obtained from Eq. (\ref{Eq_Lag_MIT1}), for two cases of non-linear interactions defined in Eqs. (\ref{Eq_Lag2_non1}) and (\ref{Eq_Lag2_non2}). 
In general, the equations of motion for $\omega, \rho$ and $\phi$ can be written as
\begin{align}
	m_\omega^2 \omega +  \frac{\partial }{\partial \cal \omega}(\mathcal{L}_{vec}^{Non})  & = g_\omega^u (\rho_u + \rho_d),
	\label{Eq_omega1} \\
	m_\rho^2 \rho + 
	\frac{\partial }{\partial \cal \rho}(\mathcal{L}_{vec}^{Non}) &= g_\rho^u (\rho_u - \rho_d), 
	\label{Eq_rho1}\\
	m_\phi^2 \phi +   
	\frac{\partial }{\partial \cal \phi}(\mathcal{L}_{vec}^{Non})  &= g_\phi^s  \rho_s,
	\label{Eq_phi1} 
\end{align}
where the derivative of $\mathcal{L}_{vec}^{Non}$ with respect to the vector fields $\omega, \rho$ and $\phi$, for two forms of non-linear interaction of vector mesons (Eqs. (\ref{Eq_Lag2_non1}) and (\ref{Eq_Lag2_non2})), are tabulated in Table \ref{Tab1_non_der1}.
The effective chemical potential, ${\mu_i}^{*}$, of quarks is defined by
\begin{equation}
	{\mu_i}^{*}=\mu_i-g_{\omega}^{i}\omega-g_{\rho}^{i}\rho-g_{\phi}^{i}\phi.
	\label{Eq_mueff}
\end{equation}

\begin{table}
	\begin{tabular}{|c|m{2cm}|m{3cm}|}
		\hline
		$\cal V$ & $\frac{\partial }{\partial \cal V}(\mathcal{L}_{vec-I}^{Non})$& 
		$\frac{\partial }{\partial \cal V}(\mathcal{L}_{vec-II}^{Non})$ \\
		\hline
		$\omega$ & $4 c_4 g^u_{\omega}$ $    \left(\omega^3 + 3 \omega \rho^2\right)$ & $4 c_4 g^u_{\omega} \omega  \left[g^u_{\omega} \left(\omega^2 + \rho^2\right)\right. $ $\left.  + g^s_{\phi} \phi^2\right]$ \\
		\hline
		$\rho$ &$4 c_4 g^u_{\omega}$ $    \left(\rho^3 + 3 \omega^2 \rho\right)$ &$4 c_4 g^u_{\omega} \rho $ $ \left[g^u_{\omega} \left(\omega^2 + \rho^2\right)\right. $ $\left.  + g^s_{\phi} \phi^2\right]$ \\
		\hline
		$\phi$ &$8 c_4 g^s_{\phi} \phi^3$ 
		& $4 c_4 g^s_{\phi} \phi $ $ \left[g^u_{\omega} \left(\omega^2 + \rho^2\right)\right. $ $\left.  + g^s_{\phi} \phi^2\right]$ \\
		\hline
	\end{tabular}
	\caption{Derivatives of non-linear interaction Lagrangian with respect to the vector mesons.} 
	\label{Tab1_non_der1}
\end{table} 
Total energy density of the cold SQS is given as
\begin{align}
	\epsilon & = \sum_{i=u,d,s,e,\mu} \gamma_i \int_{0}^{k_{F^i}} \frac{d^3k}{(2\pi)^3} \sqrt{k^2 + m_i^2}  +  \nonumber\\
	& B - \frac{1}{2}\left(m_\omega^2 \omega^2 + m_\rho^2 \rho^2 +m_\phi^2 \phi^2\right) - 
	\mathcal{L}_{vec}^{Non}.
	\label{Eq_energy_sqs1}
\end{align}
In the above equation, $k_{F}^i$ is the Fermi-momentum of particle at zero temperature and is related to corresponding particle density through relation 
\begin{align}
	\rho_i = \frac{\gamma_i k_{F}^{i3}}{6\pi^2}.
\end{align}
The parameter $\gamma_i$ is the degeneracy factor having value $\gamma_i = 6$ and $2$, for quarks and leptons, respectively.
The energy density of $\beta-$equilibrated  SQS will be calculated using Eq.(\ref{Eq_energy_sqs1})
for two different forms of non-linear interaction term, 
$\mathcal{L}_{vec}^{Non}$, of vector mesons (Eqs. (\ref{Eq_Lag2_non1}) and (\ref{Eq_Lag2_non2})).
The pressure $p$ of the medium corresponding to SQS is calculated using equation \cite{Glendenning:1997wn}
\begin{equation}
	p = -\epsilon + \sum_{i=u,d,s,e,\mu} \mu_i\rho_i.
\end{equation}
The weak $\beta-$equilibrium conditions for the strange quark stars are expressed as  
\begin{align}
	\mu_{d}=\mu_{s}=\mu_{u}+\mu_e,\quad \mu_{\mu}=\mu_e.
\end{align}
Furthermore, the condition for electric charge neutrality is written as
\begin{align}
	\frac{2}{3}\rho_u=\frac{1}{3}\rho_d+\frac{1}{3}\rho_s+\rho_e+\rho_{\mu}.
\end{align}
The total baryon density $\rho_B$ can be expressed in terms of number density of quarks as 
\begin{align}
	\rho_B=\frac{1}{3} \sum_{i = u,d,s} \rho_i.
\end{align}

\section{STRUCTURE OF SQSs}
\label{Sec_tov}
Using the EoS calculated through modified vector MIT bag model having $\omega, \rho$ and $\phi$ mesons, the mass-radius relation of SQSs can be obtained by solving the Tolman-Oppenheimer-Volkoff (TOV) equation \cite{Oppenheimer:1939ne}, written as
\begin{align}
	\frac{dP(r)}{dr}=-\frac{M(r)[\varepsilon(r)+p(r)]}{r^2}\Bigg[1+\frac{4\pi p(r)r^3}{M(r)}\Bigg]\Bigg[1-\frac{2M(r)}{r}\Bigg]^{-1},
\end{align}
here $\varepsilon(r)$ is the energy density and $p(r)$ is the pressure obtained from the EoS. $M(r)$ is the gravitational mass inside the radius $r$ of the star and is obtained by integrating the differential equation
\begin{align}
	\frac{dM(r)}{dr}=4 \pi r^2 \varepsilon(r).
\end{align}
The study of the gravitational waves from a merger event provides valuable information about the EoS profile of the dense object. It describes the relationship between pressure and density of matter in extremely compact conditions. 
Placing a massive object in the non-uniform external gravitational field of another massive object results in tidal deformations. The tidal deformability denoted as $\Lambda$, measures how an object can be distorted when subjected to an external force. A higher value of $\Lambda$ indicates that the object can be easily deformed, meaning it's not very compact. Conversely, objects with a smaller tidal deformability parameter are more deformation-resistant, implying more rigidness. It is a dimensionless quantity and is calculated by solving differential equations along with the TOV equation \cite{Hinderer:2007mb, Hinderer:2009ca}
\begin{align}
	\frac{dH(r)}{dr} &= \beta, \\
	\frac{d\beta(r)}{dr} &=  2H\left(1 - \frac{2M}{r}\right)^{-1} 
	\left\{ - 2\pi\left[5\varepsilon + 9p + \frac{d\epsilon}{dp}(\varepsilon + p)\right]  \right. \nonumber\\ & \left. + \frac{3}{r^2} + 2\left(1 - \frac{2M}{r}\right)^{-1}\left(\frac{M}{r^2} + 4\pi r p\right)^2 \right\} \nonumber \\ &+ \frac{2\beta}{r}  \left(1 - \frac{2M}{r}\right)^{-1}  \left\{-1 + \frac{M}{r} + 2\pi r^2(\varepsilon - p) \right\},
	\label{diffeq}
\end{align}
where $H(r)$ is the metric function . The integration will start from the center with the expansions $H(r ) = a_0r^2$ and $\beta(r) = 2a_0 r$ as the radius $ r\to0 $. . The tidal parameter $\Lambda$ is defined in terms of Love number $k_2$, characterizing how an extended object responds to tidal forces, through relation \cite{Hinderer:2007mb, Postnikov:2010yn}
\begin{align}
	\Lambda = \frac{2}{3}k_2C^{-5}.
\end{align}
In above, the compactness parameter $C$ quantifies the compactness of the object, and it is defined using mass $M$ and radius $R$ of the star as $C$ = $M/R$. The relationship between $k_2$ and the $C$ of an object holds great importance, as it can help us understand how the object's structure affects its reaction to tidal forces. It is given as follows\cite{Hinderer:2007mb, Postnikov:2010yn}
\begin{align}\label{lovenumc}
	k_{2}&=\frac{8}{5}C^{5}(1-2C)^{2}[2C(y_{2}-1)-y_{2}+2]\left(2C(4(y_{2}+1)C^4 \right. \nonumber\\
	&+(6y_{2}-4)C^{3}
	+(26-22y_{2})C^2+3(5y_{2}-8)C-3y_{2}+6)\nonumber\\
	&-3(1-2C)^2(2C(y{_2}-1)-y_{2}+2) \left. \log\left(\frac{1}{1-2C}\right)\right)^{-1},
\end{align}
where $y_2 = R \beta(R)/H(R) - 4\pi R^3 \epsilon_s/M$. Here, $\epsilon_s$ is the energy density at the surface of quark star. The term 
$4\pi R^3 \epsilon_s/M$ is subtracted due to non-zero value of energy density at the surface of SQSs \cite{Li:2019ztm,Lourenco:2021lpn,Albino:2021zml}.

\section{Results and discussions}
\label{sec:results}
We now present the results of our present investigation for the properties of strange quark stars. We shall calculate the EoS and properties of strange stars for $\beta-$equilibrated matter composed of $u,d$ and $s$ quarks
and leptons $e$ and $\mu$ using the vector MIT bag model considering exchange of  vector mesons $\omega$, $\rho$ and $\phi$ and considering two different cases for 
higher order non-linear self-interactions of vector mesons.
Three flavor strange matter composed of $u,d$ and $s$ quarks will be stable and true ground state of matter, if its  binding energy per baryon is 
less than 930  MeV and two flavor matter has energy per baryon above this value.
For a given value of vector coupling $g_v$, the stability condition for three flavor matter is satisfied  for a certain range of bag pressure $B$ known as stability window. 
In Table \ref{tab_Swindow}, values of $B^{1/4}$ corresponding to stability window are tabulated  
at different values of vector coupling $g_v$. In the last column average values of $B^{1/4}$ corresponding to the given stability window are tabulated. 
As one can see, with an increase in the strength of vector interactions, i.e., increasing the value of vector coupling $g_v$, the minimum and maximum values of $B^{1/4}$ shift to lower side.
For $B^{1/4}$ values less than the minimum values tabulated in Table \ref{tab_Swindow}, two flavor matter will be stable compared to three flavor strange matter. 
\begin{table}
	\begin{center}
		\begin{tabular}{|c|c|c|c|}
			\hline
			~ $g_v~$  & ~$B_{min}^{1/4}$~ & $B_{max}^{1/4}$~ & $B_{avg}^{1/4}$~ \\
			\hline
			0  &  145 & 159  & 152 \\
			1  &   143  & 157  & 150 \\
			2  &   137 & 152  & 144.5 \\
			3  &   131 & 145  & 138 \\
			\hline
		\end{tabular}
				\caption{Stability windows obtained with the vector MIT bag model of present work. All values of $B^{1/4}$ are in MeV units.} 
				\label{tab_Swindow}
			\end{center}
		\end{table}
		
\begin{figure}
			\centering
			\begin{minipage}[c]{0.23\textwidth}
				\includegraphics[width=\textwidth]{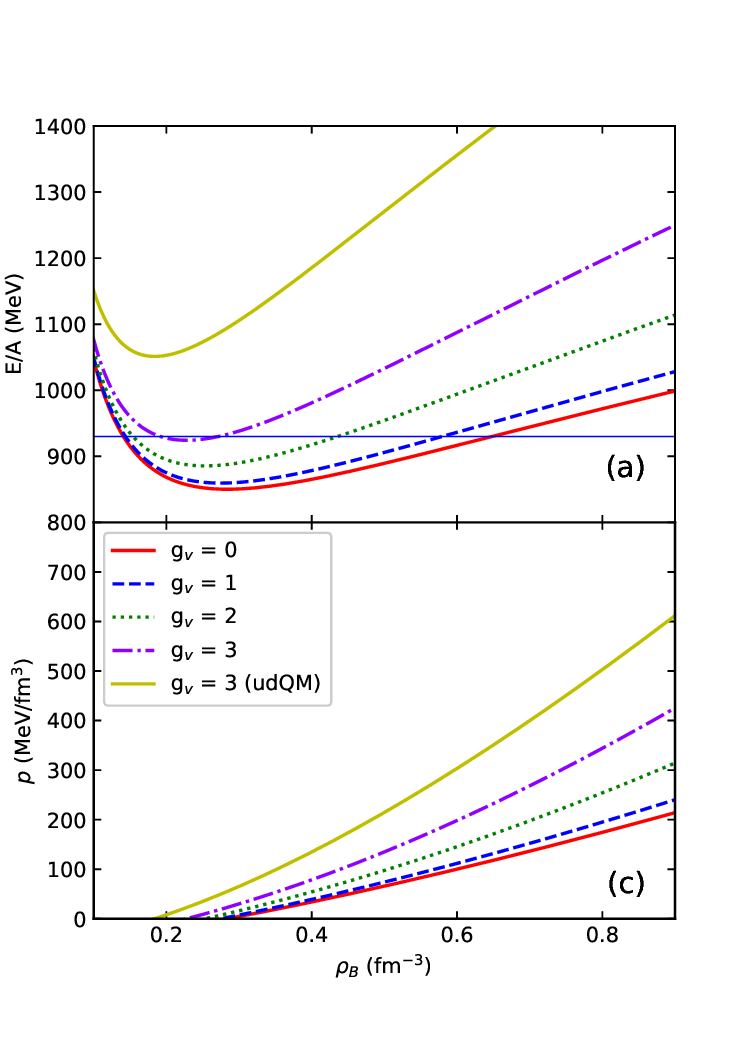}
			\end{minipage}
			\hfill
			\begin{minipage}[c]{0.23\textwidth}
				\includegraphics[width=\textwidth]{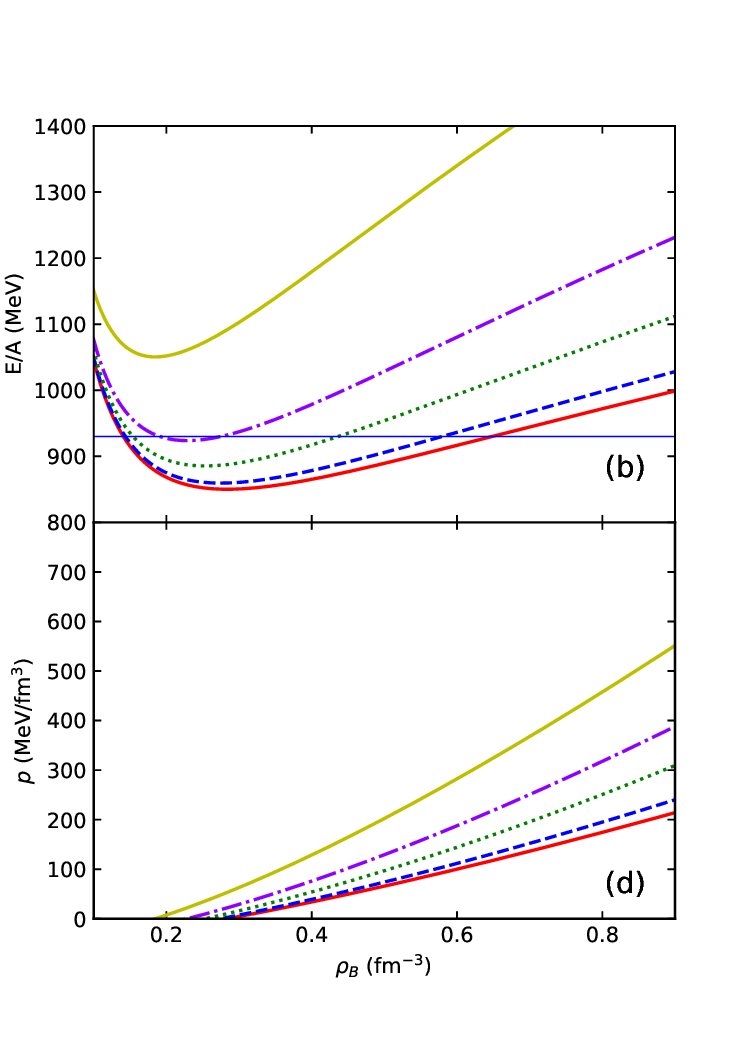}
			\end{minipage}
			\vfill
			\caption{\label{fig_BE_SQM1} (Color online)
				The energy per baryon $E/A$ is plotted as a function of baryon density $\rho_B$ in subplots (a) and (b) for $\mathcal{L}_{vec-I}^{Non}$ and $\mathcal{L}_{vec-II}^{Non}$, respectively.
				Pressure $p$ corresponding to these two cases is shown in subplots (c) and (d). In all subplots results are shown for $g_v = 0,1,2,3$ of strange matter and compared with $g_v =3$ case of two flavor matter.}
		\end{figure}

		For further confirmation of above claims regarding the stability of three flavor  matter, in subplots (a) and (b) of Fig. \ref{fig_BE_SQM1}, the energy per baryon $E/A$ of strange quark matter is plotted as a function of baryon density $\rho_B$, for two cases of non-linear interactions considered in the present work. 
		The results are plotted for the values of vector coupling $g_v = 0,1,2$ and $3$ and are compared with the $E/A$ value for two flavored quark matter composed of $u$ and $d$ quarks at $g_v =3$.
		In these results we have considered the fix value of bag parameter $B^{1/4} = 145$ MeV for all cases of $g_v$,  as this value falls in each stability window given in Table \ref{tab_Swindow}.
		As one can see, for an increase in $g_v$ of strange quark matter, the minima of curves shifts to lower values of baryon density $\rho_B$. Also, the value of $E/A$ at minima of curve increases with an increase in strength of vector coupling, indicating an increase in the strength of repulsive interactions.
		The energy per baryon of two flavored matter, composed of $u$ and $d$ quarks, shown for $g_v=3$ case is above $930$ MeV, indicating its instability  and favor of strange quark
		matter hypothesis.
		There is very small change in the behavior of curves in two cases of non-linear interactions.
		In the subplots (c) and (d) of Fig. \ref{fig_BE_SQM1} pressure $p$ is plotted as 
		a function of baryon density $\rho_B$ corresponding to $E/A$ of subplots (a) and (b), respectively.
		As expected for thermodynamic consistency of
		calculations, pressure is found to be zero at minima of $E/A$. Also, for a given $\rho_B$, the value of $p$ is observed to increase as a function of $g_v$.
		
\begin{figure}
			\begin{minipage}[c]{0.23\textwidth}
				\includegraphics[width=\textwidth]{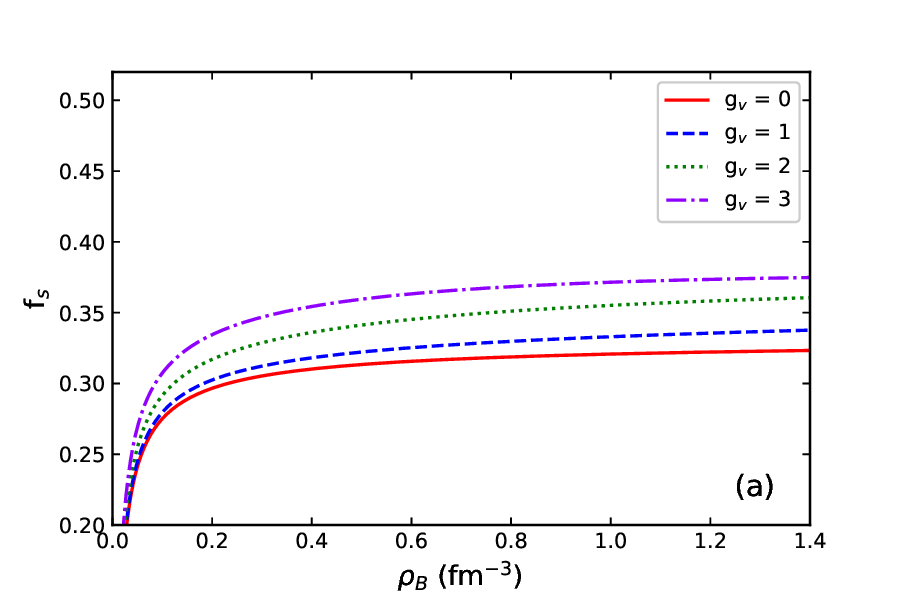}
			\end{minipage}
			\hfill
			\begin{minipage}[c]{0.23\textwidth}
				\includegraphics[width=\textwidth]{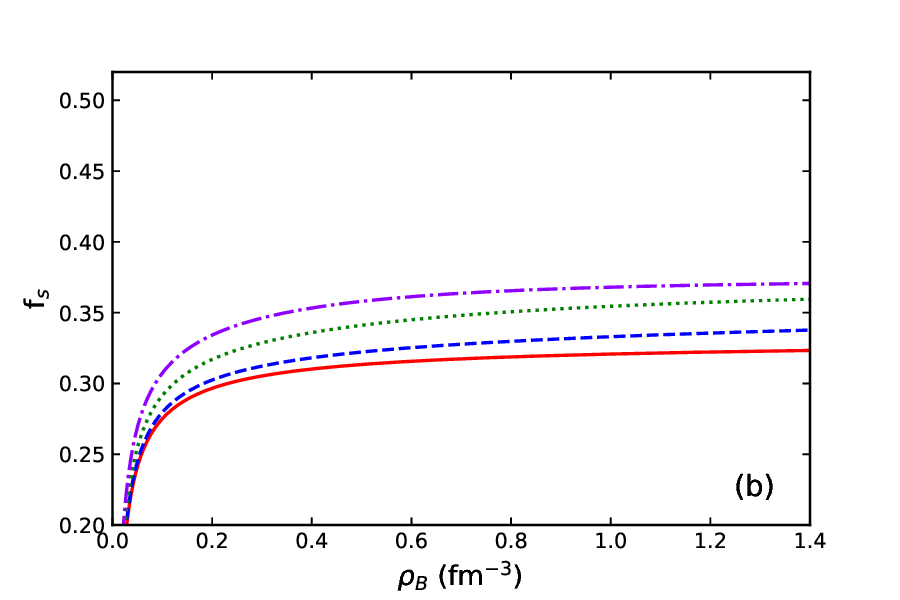}
			\end{minipage}
			\vfill
			\caption{\label{Fig_squarkf1} (Color online) In figure above, strange quark fraction $f_s$ is plotted as a function of baryon density $\rho_B$, at $B^{1/4} = 145$ MeV. 
			}
		\end{figure}

		The strange quark fraction $f_s = \rho_s/{3\rho_B}$ is plotted as a function of $\rho_B$ in Fig. \ref{Fig_squarkf1} for values of $g_v = 0,1,2$ and $3$
		and keeping $B^{1/4} =145$ MeV.
		The fraction of strange quarks in the medium is observed to increase with $g_v$. The threshold value of $\rho_B$ where strange quarks start populating also decreases to lower side with increasing $g_v$ value. 		
		
			\begin{figure}
			\centering
			\begin{minipage}[c]{0.23\textwidth}
				\includegraphics[width=\textwidth]{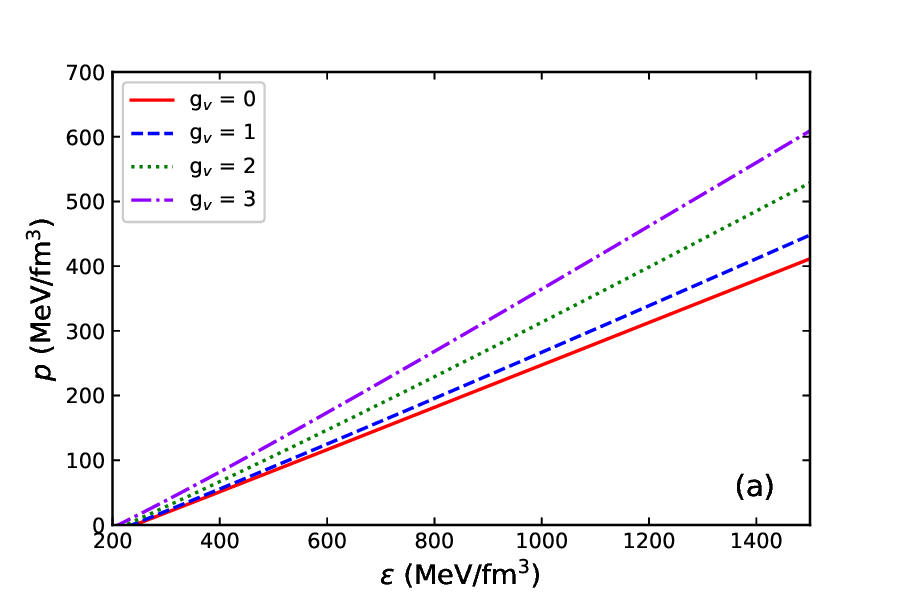}
			\end{minipage}
			\hfill
			\begin{minipage}[c]{0.23\textwidth}
				\includegraphics[width=\textwidth]{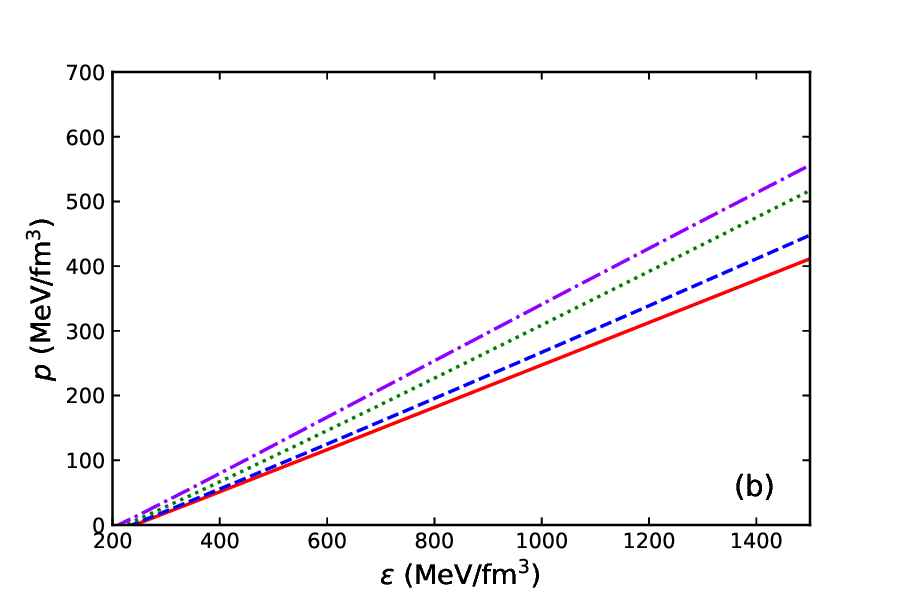}
			\end{minipage}
			
			\begin{minipage}[c]{0.23\textwidth}
				\includegraphics[width=\textwidth]{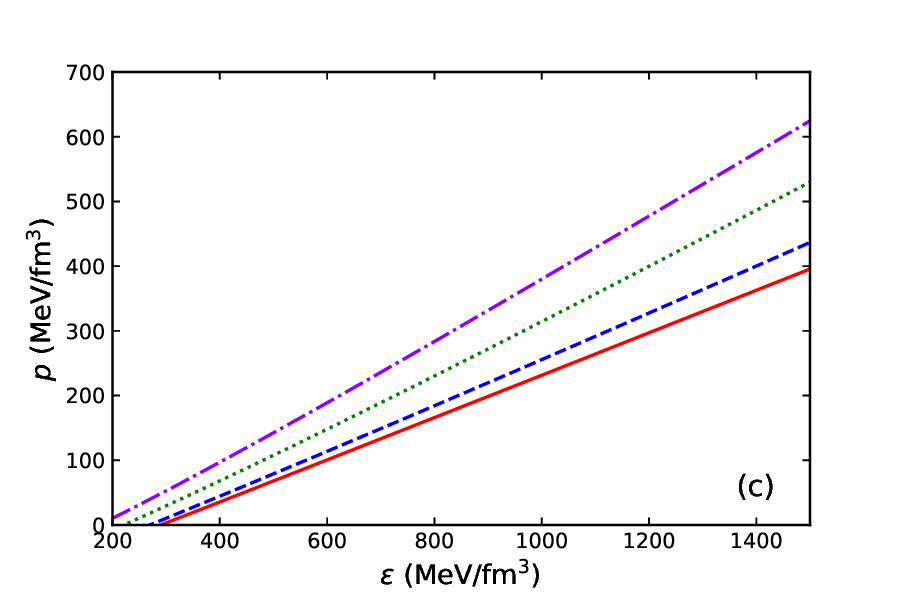}
			\end{minipage}
			\hfill
			\begin{minipage}[c]{0.23\textwidth}
				\includegraphics[width=\textwidth]{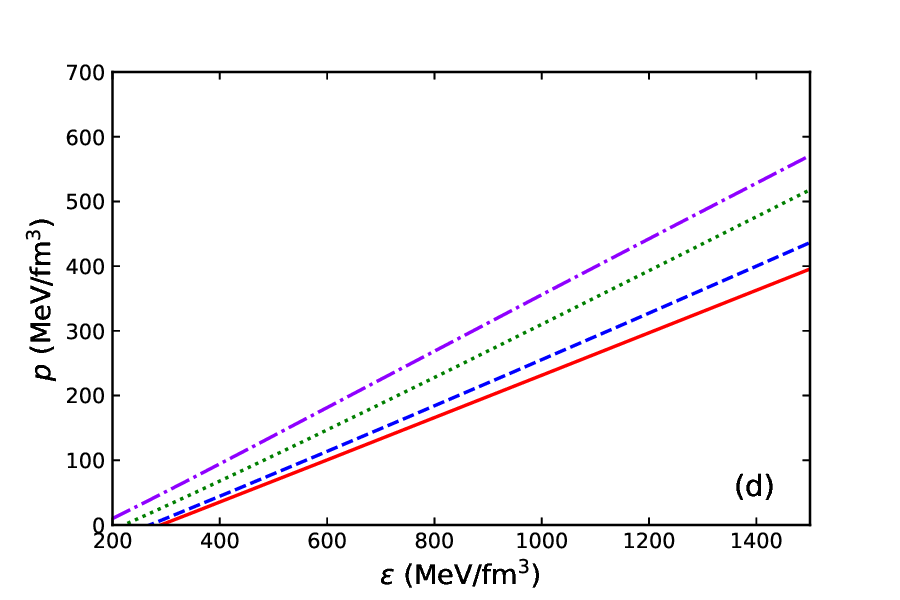}
			\end{minipage}
			\hfill
			\caption{\label{Fig_EoS1} (Color online) In the above figure the EoS of SQSs, i.e., pressure $p$ vs energy density $\epsilon$ is plotted. Subplots (a) and (b) are for
				$B^{1/4} = 145$ MeV, whereas (c) and (d) are at average value of $B^{1/4}$ given in Table \ref{tab_Swindow} for the stability window at different $g_v$ values.
				Left panel considers $\mathcal{L}_{vec-I}^{Non}$ whereas right is for $\mathcal{L}_{vec-II}^{Non}$.}
		\end{figure}
	
		In Fig. \ref{Fig_EoS1} the EoS for SQSs is plotted showing the impact of increasing strength of vector coupling on its stiffness. In subplots (a)  and (b) the results are shown for a fix value of $B^{1/4} =145$ MeV. In (c) and (d) results are shown for average value of $B^{1/4}$ calculated for each stability window as given in Table \ref{Tab1_non_der1}. Left column of figure
		depicts the results when $\mathcal{L}_{vec-I}^{Non}$ is considered
		for non-linear interactions, whereas right column is for case $\mathcal{L}_{vec-II}^{Non}$.
		For a given value of energy density $\epsilon$, an increase in the strength of vector couplings from $g_v = 0$ to $3$, result in an increase of pressure values, i.e., the EoS gets more stiffer, a favorable situation for massive compact stars.
		For a given finite value of $g_v$, EoS state is more stiff  in case of $\mathcal{L}_{vec-I}^{Non}$ compared to $\mathcal{L}_{vec-II}^{Non}$.

		\begin{figure}
			\begin{minipage}[c]{0.23\textwidth}
				\includegraphics[width=\textwidth]{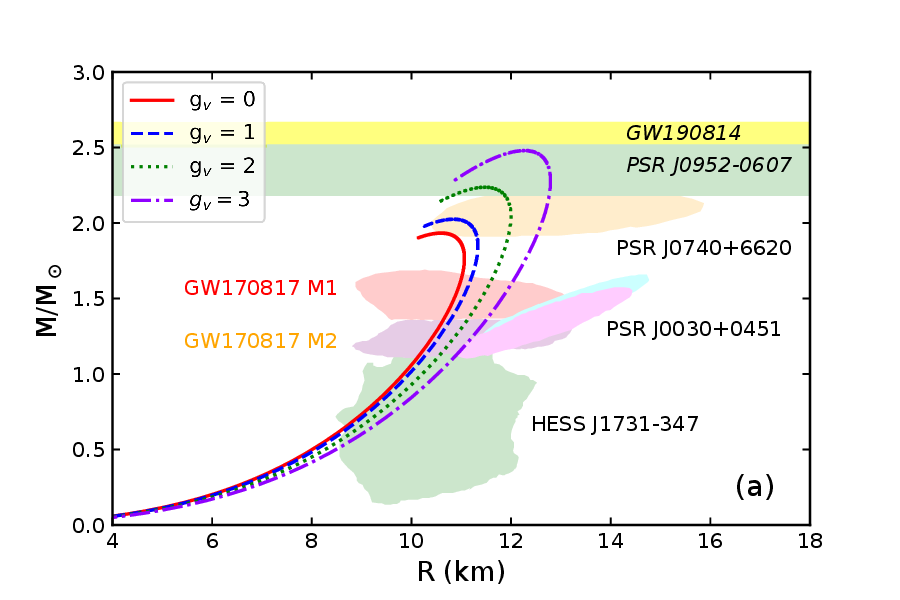}
			\end{minipage}
			\hfill
			\begin{minipage}[c]{0.23\textwidth}
				\includegraphics[width=\textwidth]{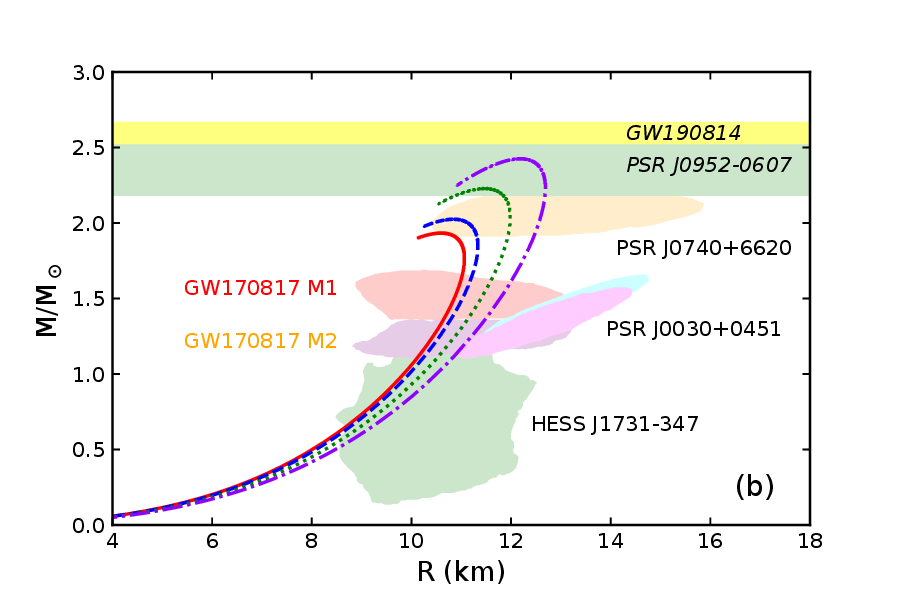}
			\end{minipage}
			
			\begin{minipage}[c]{0.23\textwidth}
				\includegraphics[width=\textwidth]{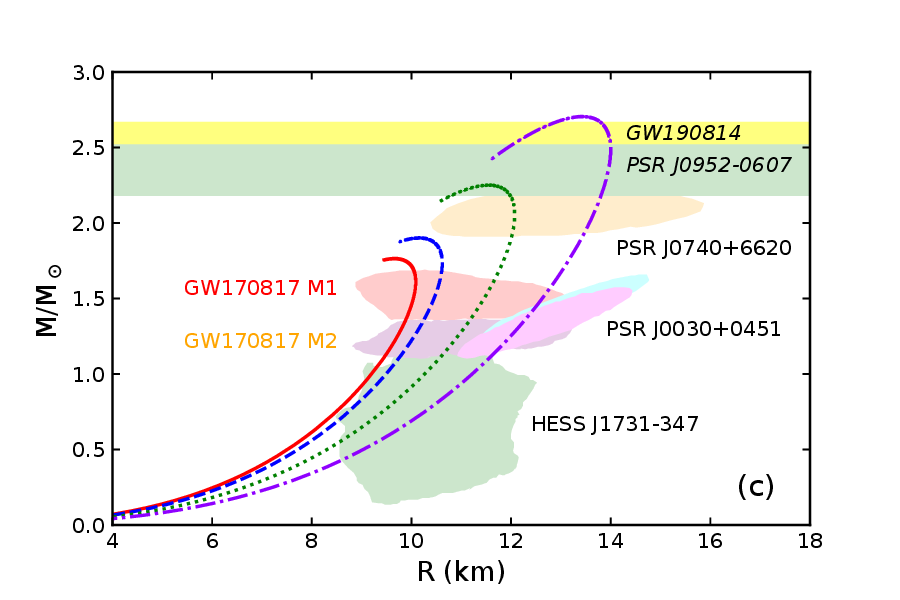}
			\end{minipage}
			\hfill
			\begin{minipage}[c]{0.23\textwidth}
				\includegraphics[width=\textwidth]{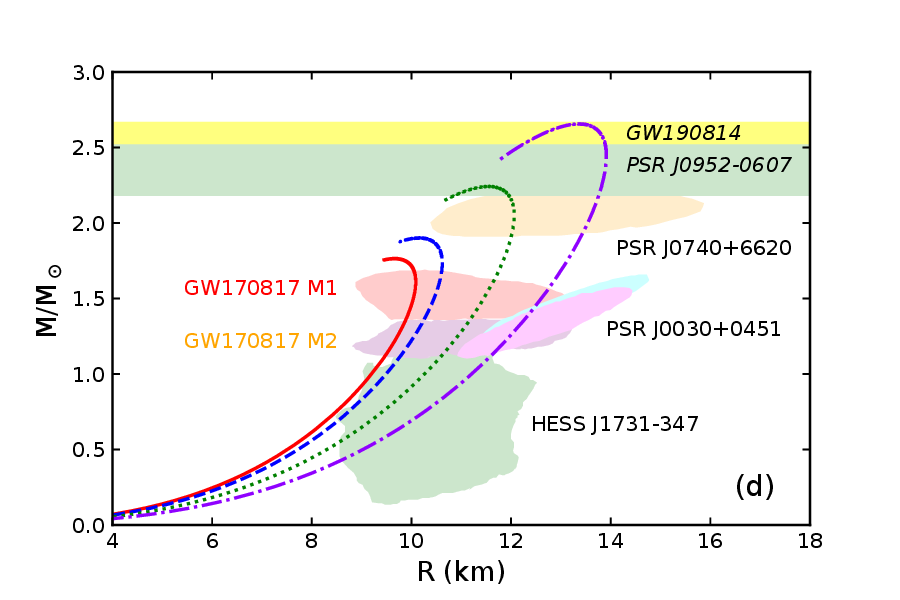}
			\end{minipage}
			\hfill
			\caption{\label{fig_MR1} (Color online) 
				In the above figure the mass-radius ($M-R$) relation is plotted for SQS, for different $g_v$ values. Description is same as in  Fig. \ref{Fig_EoS1}.
				Observational constraints on maximum mass and radius from PSR J0740+6620 \cite{Fonseca:2021wxt,Miller:2021qha}, PSR J0030+0451 \cite{Riley:2019yda,Miller:2019cac}, PSR J0952-0607 \cite{Romani:2022jhd} and HESS J1731-347 \cite{Doroshenko:2022nwp} are illustrated. The observational limits on $M-R$ plane suggested by GW170817 \cite{LIGOScientific:2018cki} and GW190814 \cite{LIGOScientific:2020zkf} are also compared.}
		\end{figure}
		
		\begin{table}
			\begin{center}
				\begin{tabular}{|c|c|m{1cm}|m{1cm}|m{1cm}|c|c|c|}
					\hline
					
					~ $g_V$ ~ & Case &$M_{max}$ ($M_\odot$) ~& R ~ (km) & R$_{1.4}$ (km) & $\Lambda_{1.4}$ \\
					\hline 
					0 & - & 1.93 & 10.60  & 10.72 & 66 \\
					\hline
					1 & \multirow{3}{*}{$\mathcal{L}_{vec-I}^{Non}$} & 2.02 & 10.82 & 10.86  & 71 \\
					2 & & 2.23 & 11.49 & 11.21  & 82 \\
					3 & & 2.48 & 12.27 & 11.60 & 94 \\ 
					\hline
					1 & \multirow{3}{*}{$\mathcal{L}_{vec-II}^{Non}$} & 2.02 & 10.82 & 10.86  & 71 \\
					2 & & 2.22 & 11.47 & 11.21  & 81 \\
					3 & & 2.42 & 12.18 & 11.57 & 93 \\
					\hline
				\end{tabular} 
				\caption{SQS properties for different values of $g_v$ at constant $B^{1/4}$}.
				\label{tab:mass_C}
			\end{center}
		\end{table}
		
		\begin{table}
			\begin{center}
				\begin{tabular}{|c|c|m{1cm}|m{1cm}|m{1cm}|c|c|c|}
					\hline
					~ $g_v$ ~ & Case & $M_{max}$ ($M_\odot$) ~& R ~ (km) & R$_{1.4}$ (km) & $\Lambda_{1.4}$ \\
					\hline 
					0 & - & 1.76 & 9.65 & 9.94 & 44 \\
					\hline
					1 & \multirow{3}{*}{$\mathcal{L}_{vec-I}^{Non}$} & 1.90 & 10.16 & 10.32  & 54 \\
					2 & & 2.25 & 11.55 & 11.28  & 82 \\
					3 & & 2.70 & 13.43 & 12.39 & 131 \\ 
					\hline
					1 & \multirow{3}{*}{$\mathcal{L}_{vec-II}^{Non}$} & 1.90 & 10.16 & 10.32 & 54 \\
					2 & & 2.24 & 11.53 & 11.26 & 84 \\
					3 & & 2.65 & 13.36 & 12.38 & 131 \\
					\hline
				\end{tabular} 
				\caption{SQS properties for different values of $g_v$ at $B^{1/4}_{avg}$.}
				\label{tab:Mass_avg}
			\end{center}
		\end{table}  
		The mass-radius relations of SQSs are obtained by solving the TOV equations as described in Sec. \ref{Sec_tov}.
		Fig. \ref{fig_MR1} shows the $M-R$ curves for SQSs for two cases of non-linear interaction $\mathcal{L}_{vec}^{Non}$. In each subfigure,
		the $M-R$ curves are plotted for $g_v = 0,1,2,3$. Similar to Fig.~\ref{Fig_EoS1},
		subplots (a) and (b) are for
		fix value $B^{1/4} = 145$ MeV, whereas (c) 
		and (d) are for average value of $B^{1/4}$ in each stability window. The SQS's mass-radius configuration is consistent with the astrophysical observation data provided by the NICER experiment for PSR J0740+6620 \cite{Fonseca:2021wxt,Miller:2021qha}, PSR J0030+0451 \cite{Riley:2019yda,Miller:2019cac}, and data from GW170817 \cite{LIGOScientific:2018cki}, GW190814 \cite{LIGOScientific:2020zkf}, PSR J0952-0607 \cite{Romani:2022jhd} and HESS J1731-347 \cite{Doroshenko:2022nwp}.
		In Table \ref{tab:mass_C} values of maximum mass $M_{max}$, corresponding radius $R$ and tidal deformability for canonical mass $\Lambda_{1.4}$ 
		are given at  $B^{1/4} = 145$ MeV.
		Values of these macroscopic properties of SQSs for average value of $B^{1/4}$ are given in Table \ref{tab:Mass_avg}.
		As one can see, without inclusion of vector mesons in the vector MIT bag model, $g_v = 0$, massive compact stars cannot be produced as desired by recent constraints from different astronomical observations.
		However, when the vector interactions are taken into account, due to stiffness of the EoS, the maximum mass is observed to increase. 
		For the vector couplings $g_v = 2$ and $3$, the value of maximum mass is found to be larger in case of $\mathcal{L}_{vec-I}^{Non}$ as compared to
		$\mathcal{L}_{vec-II}^{Non}$ due to soft EoS in the later case.
		Comparing the results of maximum mass of SQS, for $B^{1/4} = 145$ MeV (Table \ref{tab:mass_C}) and average $B^{1/4}$ (Table~\ref{tab:Mass_avg}), values are observed to be more in later case at $g_v = 2$ and $3$. 
		If the higher order non-linear interactions are not considered in the present calculations, i.e., the second term on the left hand side of Eqs. (\ref{Eq_omega1}) to (\ref{Eq_phi1}) is neglected, then for $B^{1/4} = 145$ MeV, the maximum mass of SQS  is found to be $2.02$, $2.24$, and $2.50$ $M_\odot$, at $g_v = 1, 2$ and $3$, respectively.
		In this case, corresponding values of radii are found to be $10.82$, $11.47$, and $12.31$ km, respectively.
		
\begin{figure} 
			\begin{minipage}[c]{0.49\textwidth}
				\includegraphics[width=\textwidth]{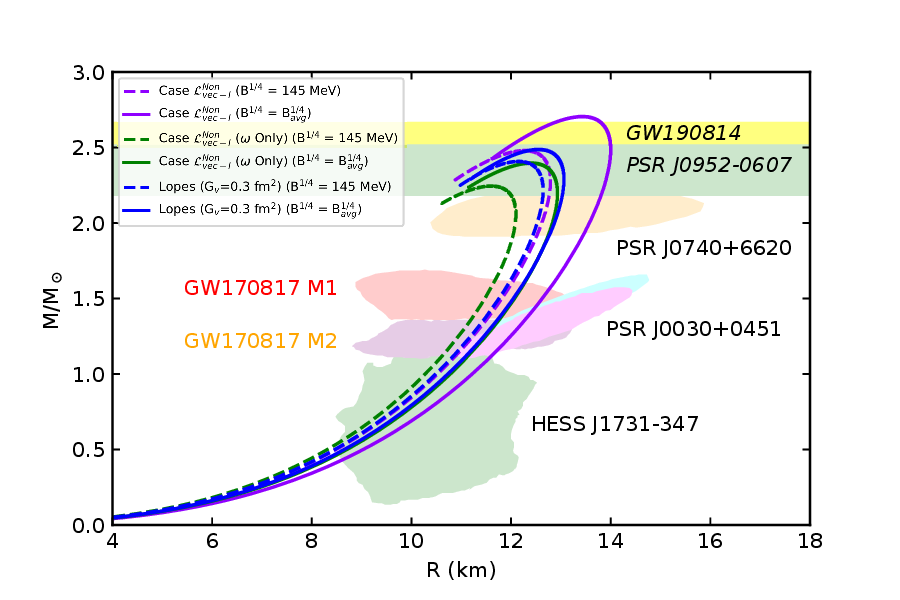}
			\end{minipage}
			\hfill
			\caption{\label{fig_MRcompare} (Color online) Comparison of mass-radius results with Ref. \cite{Lopes:2020btp}.}
		\end{figure}

		In Fig. \ref{fig_MRcompare}, we compare the mass-radius relation ($M-R$) calculated in the present work for different cases of $B^{1/4}$ at $g_v = 3$, including the case where only the $\omega$ meson is considered in our calculations, and the results of Ref. \cite{Lopes:2020btp}. When only the $\omega$ meson is considered in case $\mathcal{L}_{vec-I}^{Non}$, the stability window for $g_v = 3$ is found to range from 132 MeV to 148 MeV. Note that in Ref. \cite{Lopes:2020btp}, only $\omega$ meson was considered in the vector channel.
		Also  the coupling of $\omega$ mesons with strange $s$ quark was considered non-zero, whereas as discussed earlier, in our present calculations, coupling of $\omega$ mesons (having no $s$ quark content) with $s$ quarks is taken as zero whereas the with $\phi$ meson field
		it is considered finite. This comparison demonstrates that the inclusion of $\rho$ and $\phi$ mesons significantly enhances both the mass and radius.
		
		\begin{figure}
			\begin{minipage}[c]{0.23\textwidth}
				\includegraphics[width=\textwidth]{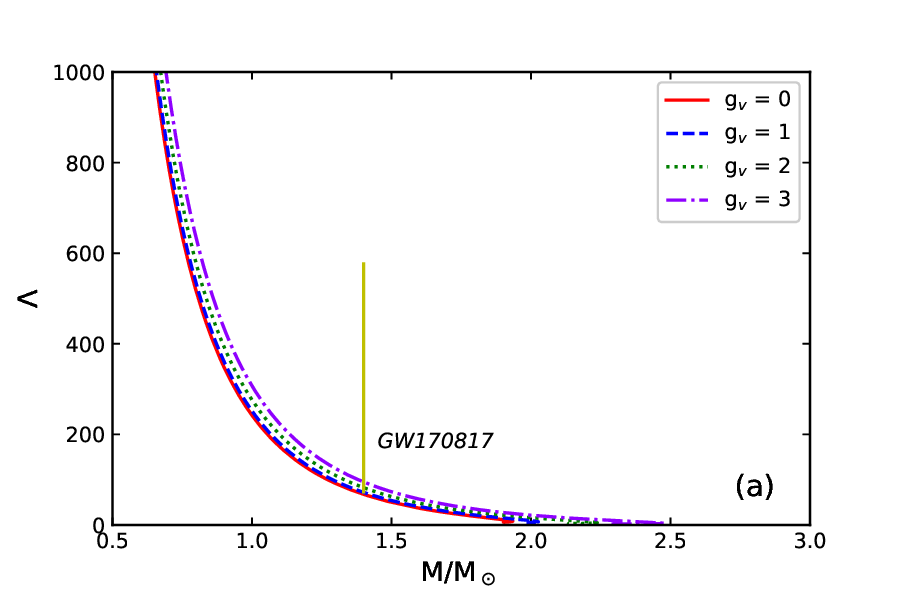}
			\end{minipage}
			\hfill
			\begin{minipage}[c]{0.23\textwidth}
				\includegraphics[width=\textwidth]{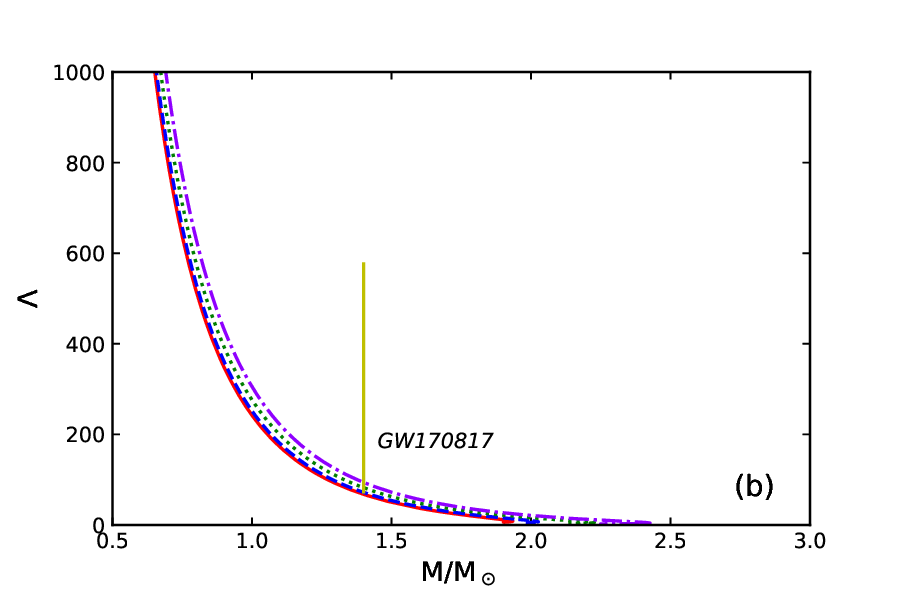}
			\end{minipage}
			\begin{minipage}[c]{0.23\textwidth}
				\includegraphics[width=\textwidth]{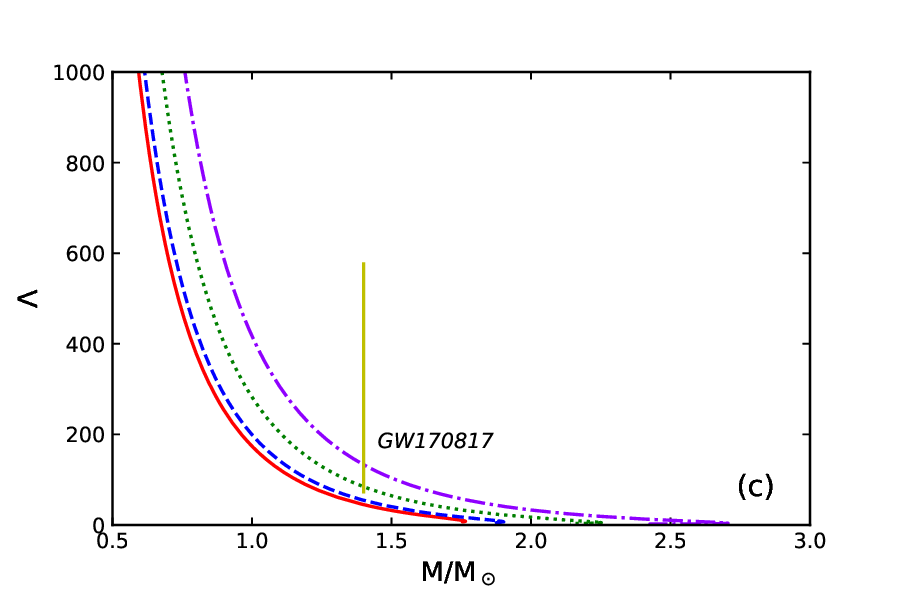}
			\end{minipage}
			\hfill
			\begin{minipage}[c]{0.23\textwidth}
				\includegraphics[width=\textwidth]{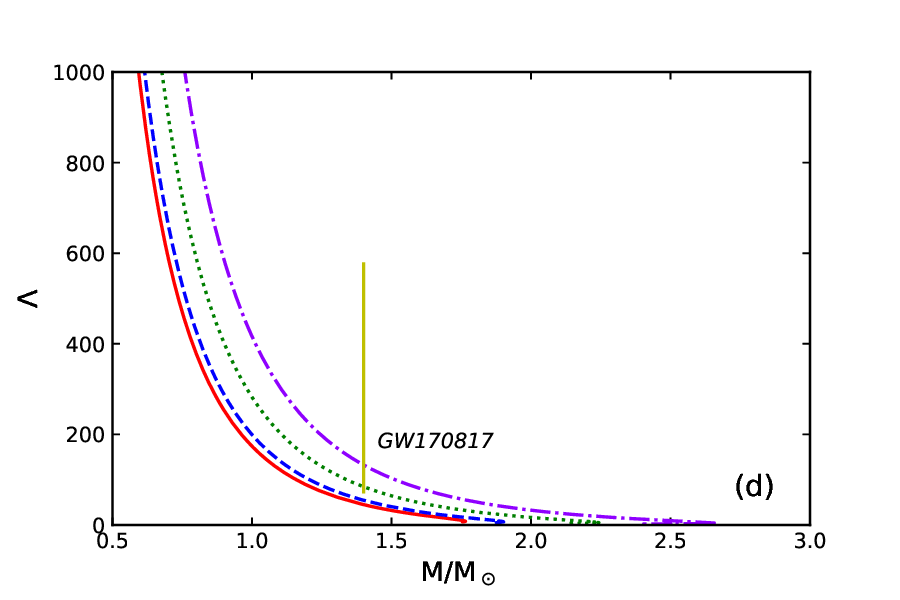}
			\end{minipage}
			\caption{\label{fig_tidal1} (Color online) In above figure, the tidal deformability parameter $\Lambda$ is plotted as a function of $M/M_{\odot}$.Description is same as in Fig. \ref{Fig_EoS1}. The vertical arrows represent the constraints on $\Lambda_{1.4}$ from GW170817 \cite{LIGOScientific:2018cki} events data. }
				\end{figure}
			
		The tidal deformability $\Lambda$ of SQS calculated using the vector MIT bag model of present work is shown in Fig. \ref{fig_tidal1} as a function of $M/M_{\odot}$. The vertical arrows represents the constraints on $\Lambda_{1.4}$ from GW170817 \cite{LIGOScientific:2018cki} events data. In the last two columns of Tables \ref{tab:mass_C}
		and \ref{tab:Mass_avg} we have given the values of radius and
		tidal deformability of canonical mass star, i.e., stars having mass $M = 1.4 M_{\odot}$.
		The tidal deformability parameter $\Lambda$ is observed to increase with increase in the value of coupling constant $g_v$.
		As can be seen from Fig. \ref{fig_tidal1}, increase in the value of 
		$\Lambda$ is more appreciable for low mass stars. Also, for  $g_v = 2$ and $3$, comparing the results for different forms of non-linear interactions, the tidal deformabilities are smaller in stars having softer EoS.  This observation is also consistent when results of 
		Tables \ref{tab:mass_C}
		and \ref{tab:Mass_avg} are compared at $g_v = 2$ and 3.
		The constraint $70 \leq \Lambda_{1.4} \leq 580$ imposed by GW170817 is seen to be satisfied in our calculations at $B^{1/4}=145$ MeV for values of $g_v=1,2,3$ and for the average value of $B^{1/4}$ at $g_v=2,3$.

	\begin{figure}
	\begin{minipage}[c]{0.23\textwidth}
		\includegraphics[width=\textwidth]{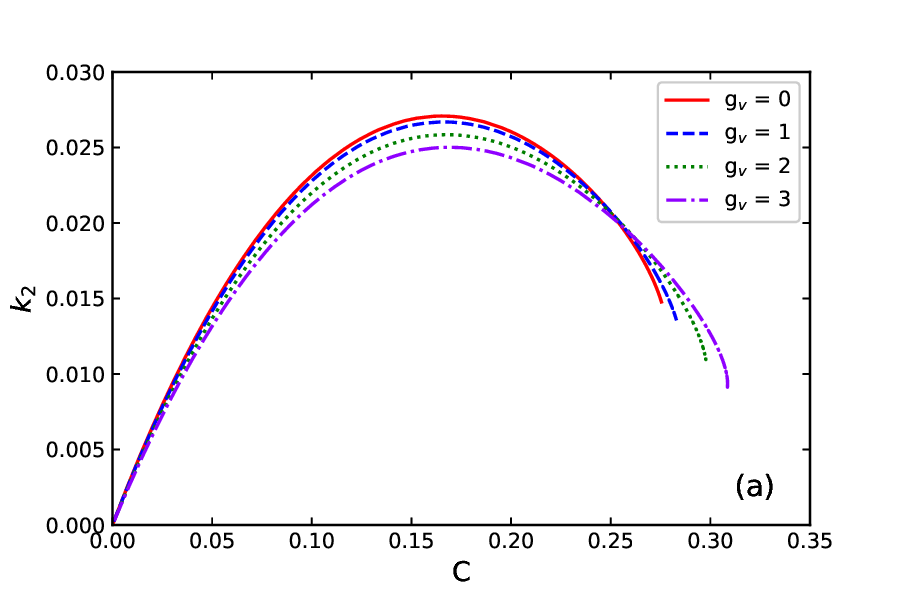}
	\end{minipage}
	\hfill
	\begin{minipage}[c]{0.23\textwidth}
		\includegraphics[width=\textwidth]{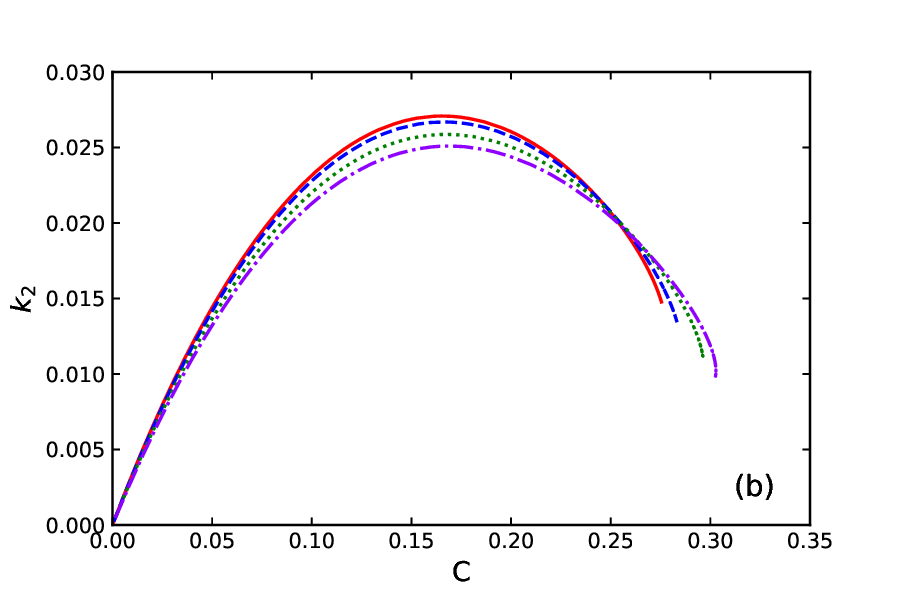}
	\end{minipage}
	\begin{minipage}[c]{0.23\textwidth}
		\includegraphics[width=\textwidth]{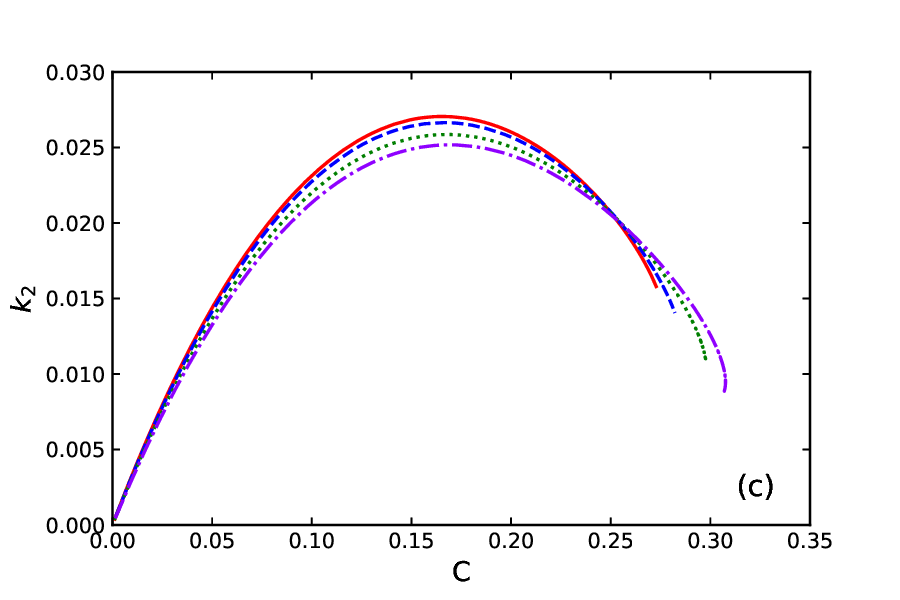}
	\end{minipage}
	\hfill
	\begin{minipage}[c]{0.23\textwidth}
		\includegraphics[width=\textwidth]{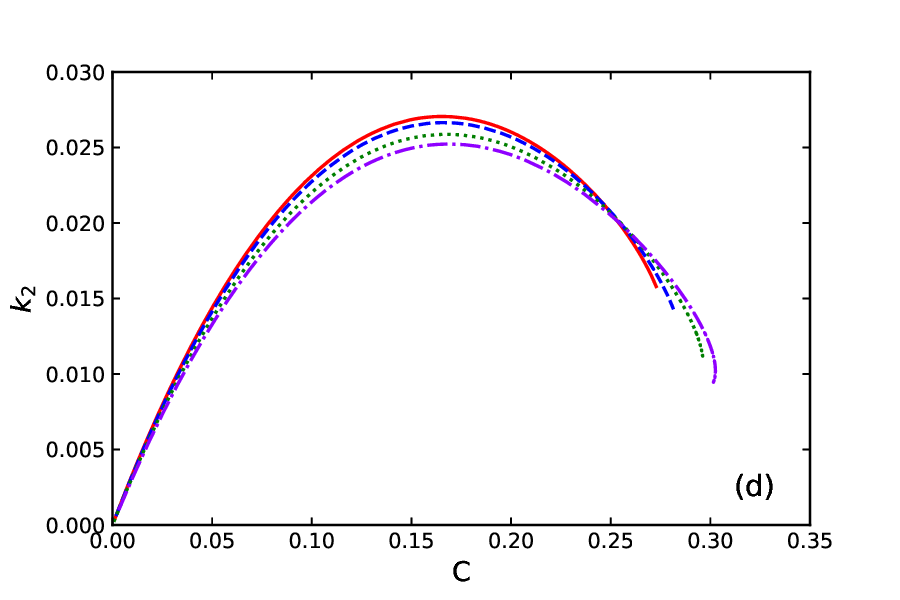}
	\end{minipage}
	
	\caption{\label{fig_lovek2} (Color online) In the above figure the Love number $k_2$   is plotted for SQS as a function of compactness parameter $C$, for different $g_v$ values. Description is same as
		in Fig. \ref{Fig_EoS1}.}
\end{figure}

		The Love number $k_2$ is depicted as a function of compactness parameter $C$ in Fig. \ref{fig_lovek2}. As C increases, k2 increases to a maximum of around 0.2501-0.2708, and decreases rapidly. A lower value of Love number $k_2$  is observed with increasing the strength of vector interactions and attained a maximum at a higher value of $C$. This implies that the compact stars with stiffed EoS will have a lower value of $k_2$.This observation is also consistent as one compares the results of $\mathcal{L}_{vec-I}^{Non}$ and $\mathcal{L}_{vec-II}^{Non}$.
		
\begin{figure}
			\begin{minipage}[c]{0.23\textwidth}
				\includegraphics[width=\textwidth]{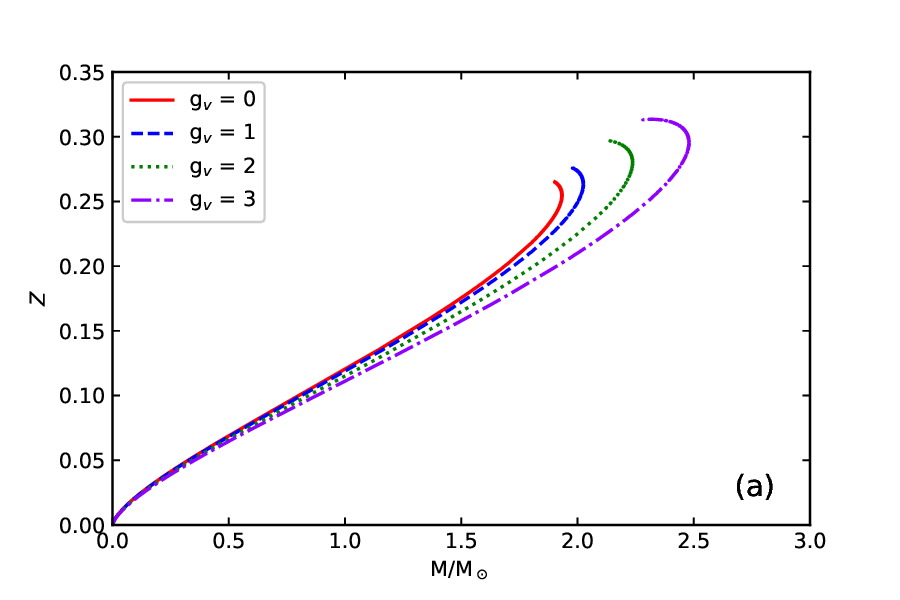}
			\end{minipage}
			\hfill
			\begin{minipage}[c]{0.23\textwidth}
				\includegraphics[width=\textwidth]{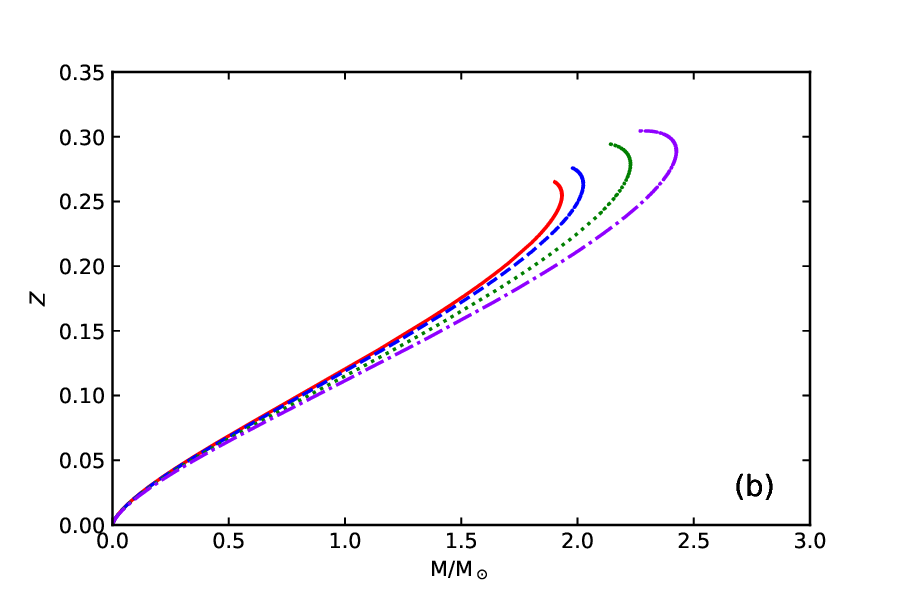}
			\end{minipage}
			\begin{minipage}[c]{0.23\textwidth}
				\includegraphics[width=\textwidth]{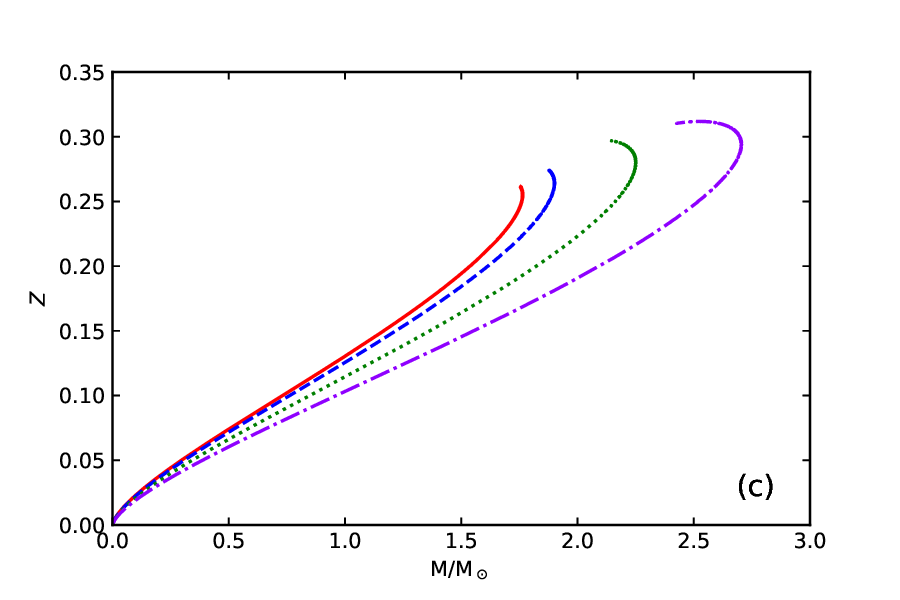}
			\end{minipage}
			\hfill
			\begin{minipage}[c]{0.23\textwidth}
				\includegraphics[width=\textwidth]{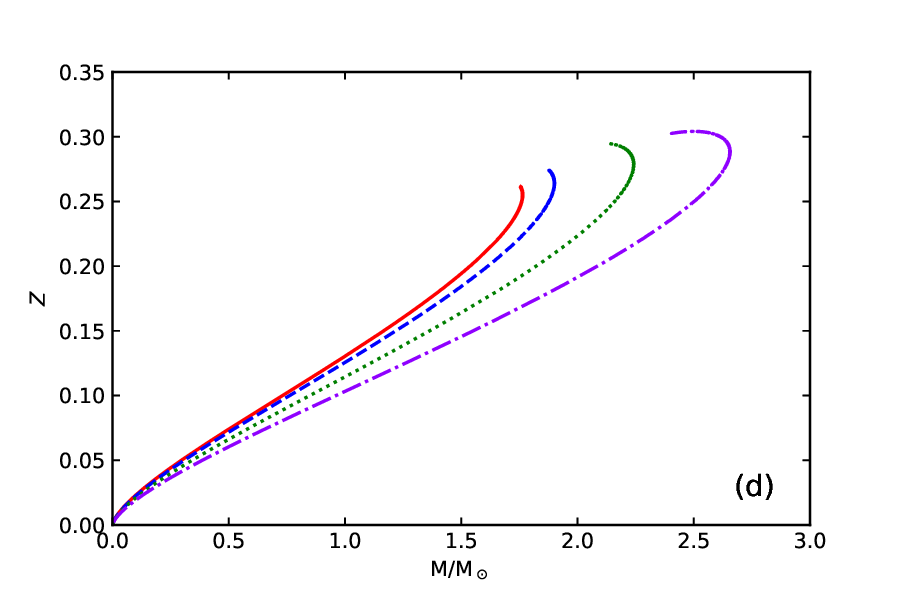}
			\end{minipage}
			
			\caption{ \label{fig_zshift1} (Color online) In the above figure the gravitation redshift $Z$ is plotted as a function of $M/M_{\odot}$ for SQS, for different $g_v$ values. Description is same as
				in Fig. \ref{Fig_EoS1}.}
		\end{figure}

		The gravitational redshift $Z$ calculated using the relation
		\begin{equation}
			Z = \left(1 - \frac{2M}{Rc^2}\right)^{-1/2} - 1, 
		\end{equation}
		is plotted in Fig. \ref{fig_zshift1} as a function of ratio $M/M_{\odot}$.
		For a given value of $g_v$, 
		values of $Z$ are observed to increase with an increase in the mass $M$ of SQSs reaching a maximum value and then start decreasing. As a function of $g_v$, the change in $Z$ is observed to be larger in SQS with stiffed EoS.
		The gravitational redshift has a larger value in case of $\mathcal{L}_{vec-I}^{Non}$ as compared to $\mathcal{L}_{vec-II}^{Non}$ for both values of $B^{1/4}$. The maximum value of the $Z$ for SQS is observed as 0.313 at $g_v=3$ with $B^{1/4}=145$ MeV and minimum value as 0.261 at $g_v=0$ with $B^{1/4}=145$ MeV for case $\mathcal{L}_{vec-II}^{Non}$. Our results of gravitational redshift lie in the observed range of strange  quark star candidate RXJ185635-3750 ($Z=0.35\pm0.15$) \cite{Prakash:2002xx}. The behavior of $Z$ with entropy and temperature is also studied in the MIT bag model (constant and density-dependent bag constant) \cite{Bordbar:2020fqj} and NJL model, respectively. It is found that the $Z$ increases with an increase in temperature and entropy. In Ref.\cite{Kumari:2021tik}, authors have also studied the effect of vector interaction on the gravitational redshift of PQS using the Polyakov Chiral Quark Mean Field model and observed increment in $Z$ with $g_v$.
		
		\section{Summary and conclusion}
		\label{sec:summary}
To summarize, in the present work we studied the properties of SQSs using the vector MIT bag model extended to include the contributions of vector mesons $\rho$ and $\phi$.		
		Along with mass terms for above vector mesons, two forms of higher order self-interactions were taken into account.
		We considered finite value for the vector coupling  of non-strange vector mesons $\omega$ and $\rho$  with light $u$ and $d$ quarks only, whereas, for vector $\phi$ mesons coupling with only strange $s$ quark is taken as finite.
		The fraction of strange quarks in the medium increases with the coupling constant $g_v$, leading to stiffer EoS, particularly for higher values of $g_v$. The stiffening effect is more pronounced in the $\mathcal{L}_{\text{vec-I}}^{\text{Non}}$ model compared to $\mathcal{L}_{\text{vec-II}}^{\text{Non}}$, allowing for the formation of massive compact stars consistent with recent astrophysical observations. We computed the $M-R$ relations of SQSs for the value of bag parameter (i) $B^{1/4} = 145$ MeV which is common for the stability windows at all values of $g_v$ considered in the present work, (ii) and also for $B^{1/4}$ which are average value in each stability window for given $g_v$. The mass-radius configurations obtained align with the constraints provided by NICER for PSR J0740+6620 and PSR J0030+0451, as well as the gravitational wave events GW170817 and GW190814, among others. The inclusion of $\rho$ and $\phi$ mesons significantly enhances both the mass and radius of SQSs, compared to models considering only the $\omega$ meson, as reflected in the stability window for $g_v = 3$. The tidal deformability parameter $\Lambda$ also increases with $g_v$, with a more appreciable effect observed in low-mass stars. For $g_v = 1, 2, 3$, our results satisfy the GW170817 constraint on $\Lambda_{1.4}$ for specific values of $B^{1/4}$, further validating the model. The gravitational redshift results are consistent with observations of the strange quark star candidate RXJ185635-3750.
		Overall, this study highlights the critical role of $\rho$ and $\phi$ mesons and vector interactions in the EoS of strange quark matter and their significance in explaining the observed properties of compact stars, paving the way for further explorations in this domain.

		\section*{Acknowledgment}
		Arvind Kumar sincerely acknowledges Anusandhan National Research Foundation (ANRF), Government of India for funding the research project under the Science and Engineering Research Board-Core Research Grant (SERB-CRG) scheme (File No. CRG/2023/000557).
		%
		%
		
		\bibliographystyle{elsarticle-num}
		
		\bibliography{ref1.bib}

	\end{document}